\documentclass{article}
\usepackage{spconf,amsmath,graphicx}
\usepackage[hidelinks]{hyperref}
\usepackage{multirow}

\usepackage{amsmath,amssymb,amsfonts}
\usepackage{bm}
\usepackage{graphicx}
\usepackage{microtype}
\usepackage{textcomp}
\usepackage{booktabs}
\usepackage{algorithm}
\usepackage[noend]{algpseudocode}
\algrenewcommand{\algorithmiccomment}[1]{\hfill{\footnotesize$\triangleright$~#1}}
\usepackage{xcolor}
\usepackage{orcidlink}
\def\BibTeX{{\rm B\kern-.05em{\sc i\kern-.025em b}\kern-.08em
    T\kern-.1667em\lower.7ex\hbox{E}\kern-.125emX}}

\usepackage{tikz}
\usetikzlibrary{arrows.meta,positioning,calc,fit,backgrounds,decorations.pathreplacing}

\definecolor{mixc}{HTML}{0F7A8C}   \definecolor{resc}{HTML}{B04A6F}   \definecolor{stemc}{HTML}{2F7A4F}  \definecolor{inkc}{HTML}{131A24}

\usepackage[colorinlistoftodos]{todonotes}

\usepackage[nolist,nohyperlinks]{acronym}

\usepackage{cleveref}

\crefname{figure}{Fig.}{Figs.}
\Crefname{figure}{Fig.}{Figs.}

\crefname{section}{Sec.}{Secs.}
\Crefname{section}{Section}{Sections}

\crefname{subsection}{Sec.}{Secs.}
\Crefname{subsection}{Section}{Sections}

\usepackage{siunitx}

\renewcommand{\vec}[1]{\mathbf{#1}}

\makeatletter
\g@addto@macro\small{\setlength\abovedisplayskip{3pt plus 2pt minus 1pt}\setlength\belowdisplayskip{3pt plus 2pt minus 1pt}}
  \makeatother
\begin{document}
\ninept

\begin{acronym}
\acro{mss}[MSS]{music source separation}
\acro{vdbo}[VDBO]{vocals, drums, bass, and other}
\acro{bs}[BS]{band-split}
\acro{tf}[TF]{time-frequency}
\acro{mhsa}[MHSA]{multi-head self-attention}
\acro{pca}[PCA]{principal component analysis}
\acro{ffn}[FFN]{feed-forward network}
\acro{swiglu}[SwiGLU]{swish gated linear unit}
\acro{sota}[SOTA]{state of the art}
\acro{seg}[SEG]{stem embedding generator}
\acro{ode}[ODE]{Ordinary Differential Equation}
\acro{gt}[GT]{ground truth}
\acro{vdbgpo}[VDBGPO]{vocals, drums, bass, guitar, piano, and other}
\acro{vdbgp}[VDBGP]{vocals, drum, bass, guitar, and piano}
\acro{or-pit}[OR-PIT]{one-and-rest permutation invariant training}
\acro{pit}[PIT]{permutation invariant training}
\acro{llm}[LLM]{large language model}
\acro{cfm}[CFM]{conditional flow matching}
\acro{ms3d}[MuS3D]{Music Source Separation via Stem Discovery}
\acro{dit}[DiT]{diffusion transformer}
\acro{snr}[SNR]{signal-to-noise ratio}
\acro{ditsep}[DiT-Sep]{diffusion transformer separator}
\end{acronym}

\title{
Music Source Separation via Stem Discovery
}

\name{
    V. Valtteri Kallinen\orcidlink{0000-0002-7171-2553} \qquad 
    Eloi Moliner\orcidlink{0000-0001-5719-326X} \qquad
    Lauri Juvela\orcidlink{0000-0002-2201-103X} \qquad  
    Vesa Välimäki\orcidlink{0000-0002-7869-292X}
}

\address{Acoustics Lab, Dept.~of Information and Communications Eng., Aalto University, Espoo, Finland}

\maketitle

\begin{abstract}
\Ac{mss} methods aim to extract stems from music mixtures, which is important, for example, in karaoke, music remixing, and pedagogical applications.
While earlier research on \ac{mss} systems has been dominated by models targeting narrow sets of general stems, there have recently been attempts to support broader source definitions.
One of these methods uses audio queries to provide direct and descriptive control over the desired separation targets based on the sound itself.
However, query-based separation remains cumbersome due to the need to provide audio examples with features matching the sources contained within mixtures.
This paper proposes \ac{ms3d}, a query-based source separation framework that iteratively discovers active sources from the mixture.
On correctly detected sources, our model matches manually queried baselines and surpasses state-of-the-art text-based models.
Subjective evaluation indicates encoding artifacts as the limiting factor in current generative separation.
The findings suggest that audio-based query representations offer an effective and automatable interface for source separation.
\end{abstract}

\acresetall


\section{Introduction}

\Ac{mss} refers to the task of dividing music mixtures into components that represent individual elements of the musical mix, such as vocals, percussion, or accompaniment instruments \cite{cano2018mss}.
It is relevant to practical use cases such as karaoke track generation, music remixing and sampling, and pedagogy.

The dominant line of research on \ac{mss} has been dedicated to static settings with targets explicitly defined by the datasets as exemplified by recent research challenges \cite{mitsufuji:2022:music,fabbro:2024:sound} with the goal of splitting the mix into four components: \ac{vdbo}.
While a few systems have extended the static paradigm to additional instruments such as piano or guitar \cite{hennequin2020spleeter, rouard2023hybrid}, this rigid approach ultimately fails to scale. 
Many sound sources do not map cleanly onto a fixed instrument taxonomy, and naively growing the output vocabulary introduces severe target sparsity, since most instruments would be absent from any given mixture.

Recent studies have proposed models that offer the capability of separating arbitrary stems not strictly defined by the training data post hoc.
Text-prompt-based systems rely on an additional text query that describes the sounds to be separated from the mixture \cite{liu:2022:separate, liu:2025:separate}.
Similarly, audio queries provide a more semantically specific method of describing the desired target sounds \cite{chen2022zero,watcharasupat:2024:stem-agnostic,wen2026promptsep}.
Hyperbolic and hyperellipsoid queries instead provide region queries in the embedding space \cite{petermann:2023:hyperbolic, watcharasupat:2026:hyperellipsoidal}, while multi-query systems combine modalities such as image, text, or sound \cite{cheng:2025:omnisep, shi:2025:sam}.
However, these methods require suitable manual queries, which are difficult to provide even with knowledge about the ground truth sources, limiting their use to, e.g., large-scale automated \ac{mss}.

We denote systems that isolate all constituent sources in a mixture without prior knowledge of their identities or count as \emph{blind}.
Alternative blind approaches either use a predetermined number of outputs \cite{reddy2023audioslots},
inherit the limitations of static output vocabularies \cite{takahashi:2019:recursive}, or use a combination of \acp{llm} to describe the sources in the mixture, requiring accurate textual descriptions \cite{mahmud:2024:opensep}.
While these methods were developed for speech or general sound scenes, no prior work addresses blind separation for music, where sources exhibit strong temporal and harmonic correlations.

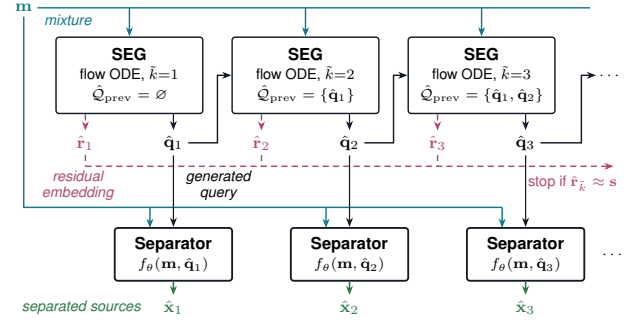
\begin{figure}[t]
    \centering
\resizebox{\columnwidth}{!}{\definecolor{mixc}{HTML}{0F7A8C}    \definecolor{resc}{HTML}{B04A6F}    \definecolor{stemc}{HTML}{2F7A4F}   \definecolor{inkc}{HTML}{131A24}

\resizebox{\columnwidth}{!}{\begin{tikzpicture}[
  font=\sffamily\footnotesize,
  >={Stealth[length=1.6mm,width=1.2mm]},
  segbox/.style  = {draw=inkc, line width=.7pt, rounded corners=1.5pt,
                    minimum width=25mm, minimum height=13mm, align=center,
                    inner sep=1.5pt},
  sepbox/.style  = {draw=inkc, line width=.9pt, rounded corners=1.5pt,
                    minimum width=20mm, minimum height=9mm, align=center,
                    inner sep=1.5pt},
  sig/.style     = {inner sep=1.5pt, outer sep=1pt},
  wire/.style    = {draw=inkc, line width=.6pt, ->},
  mixwire/.style = {draw=mixc, line width=.7pt, ->},
  reswire/.style = {draw=resc, line width=.7pt, densely dashed, ->},
  qwire/.style   = {draw=inkc, line width=.7pt, ->},
  lbl/.style     = {font=\sffamily\scriptsize, inner sep=1.5pt},
  gloss/.style   = {font=\sffamily\scriptsize\itshape, inner sep=1.5pt,
                    anchor=west},
]

\node[sig, text=mixc]   (mix)  at (0.15,0) {$\mathbf{m}$};
\node[gloss, text=mixc, anchor=south] at (0.9,-0.35) {mixture};
\draw[draw=mixc, line width=.7pt] (mix.east) -- (9.80,0);

\foreach \i/\x/\prev in {1/1.95/{$\hat{\mathcal{Q}}_{\mathrm{prev}}=\varnothing$},
                         2/4.95/{$\hat{\mathcal{Q}}_{\mathrm{prev}}=\{\hat{\mathbf{q}}_{1}\}$},
                         3/7.95/{$\hat{\mathcal{Q}}_{\mathrm{prev}}=\{\hat{\mathbf{q}}_{1},\hat{\mathbf{q}}_{2}\}$}}{
  \node[segbox] (s\i) at (\x,-1.15) {\textbf{SEG} \\[.5pt]
    {\scriptsize  flow ODE, $\tilde k{=}\i$}
    \\[.5pt]
    {\scriptsize \prev}};
  \draw[mixwire] (\x,0) -- (s\i.north);
  \node[sig, text=resc] (r\i) at (\x-0.75,-2.3) {$\hat{\mathbf{r}}_{\i}$};
  \node[sig]            (q\i) at (\x+0.75,-2.3) {$\hat{\mathbf{q}}_{\i}$};
  \draw[reswire] (r\i |- s\i.south) -- (r\i.north);
  \draw[wire]    (q\i |- s\i.south) -- (q\i.north);
}

\draw[qwire] (q1.east) -- (3.45,-2.3) -- (3.45,-1.15) -- (s2.west);
\draw[qwire] (q2.east) -- (6.45,-2.3) -- (6.45,-1.15) -- (s3.west);
\draw[qwire] (q3.east) -- (9.65,-2.3) -- (9.65,-1.15) -- (9.90,-1.15);
\node[lbl] at (10.15,-1.15) {$\cdots$};

\coordinate (lane) at (0,-2.7);
\foreach \i in {1,2,3}{\draw[draw=resc, line width=.7pt, densely dashed]
       (r\i.south) -- (r\i |- lane);}
\draw[reswire] (r1 |- lane) -- (10.20,-2.7);
\node[lbl, text=resc, anchor=north] at (9.5,-2.8)
     {stop if $\hat{\mathbf{r}}_{\tilde k}\approx\mathbf{s}$};

\node[gloss, text=resc, anchor=north west, align=center] at (0.40,-2.72)
     {residual\\embedding};
\node[gloss, anchor=north west, align=center]            at (2.85,-2.72)
     {generated\\query};

\draw[draw=mixc, line width=.7pt] (mix.south) |- (8.30,-3.4);
\foreach \i/\x in {1/2.7, 2/5.70, 3/8.70}{
  \node[sepbox] (f\i) at (\x,-4.2) {\textbf{Separator}\\[.5pt]
    {\scriptsize $f_\theta(\mathbf{m},\hat{\mathbf{q}}_{\i})$}};
  \draw[wire]    (q\i.south) -- (f\i.north);
  \draw[mixwire] (\x-0.4,-3.4) -- (\x-0.4,-3.75);
  \node[sig, text=stemc] (x\i) at (\x,-5.1) {$\hat{\mathbf{x}}_{\i}$};
  \draw[draw=stemc, line width=.8pt, ->] (f\i.south) -- (x\i.north);
}
\node[lbl] at (10.20,-4.2) {$\cdots$};
\node[gloss, text=stemc, anchor=east] at (2.2,-5.1) {separated sources};

\end{tikzpicture}}
}
\vspace{-20pt}
    \caption{Block diagram of the proposed \ac{ms3d} framework for the first three iterations, each of which separates one source or stem group.
    \vspace{-5pt}
    }
    \label{fig:diagram}
\end{figure}

Motivated by these shortcomings, this paper proposes \ac{ms3d}\footnote{Pronounced like \emph{mused}.}. 
As illustrated in \Cref{fig:diagram}, \ac{ms3d} uses \iac{seg}, also introduced in this work, to generate a set of embeddings for the constituent stems in the music mixture.
These embeddings then serve as conditioning queries for a source separator, enabling it to extract each target sound from the mix.
Experiments demonstrate that \ac{ms3d} separates music mixtures into fine-grained instrument stems without requiring external side information or manual queries.

The rest of the paper is organized as follows.
\Cref{sec:methods} describes the \ac{ms3d} framework. \Cref{sec:experiments} details the experimental setup. \Cref{sec:evaluation} outlines the results from objective metric and subjective listening evaluations.
\Cref{sec:conlusion} concludes the paper.
Sound examples for separated stems are available online.\footnote{http://research.spa.aalto.fi/publications/papers/mus3d}

\section{Methods}\label{sec:methods}

Formally, source separation describes the task of decomposing a
mixture $\mathbf{m}= \sum_{k=1}^K \mathbf{x}_k$ into its $K$ underlying components $\mathbf{x}_k \in \mathbb{R}^{C\times T}$, where $C$ denotes the number of channels and $T$ the number of samples.
In \ac{mss}, the components $\vec{x}_k$ generally consist of vocal or instrumental tracks, or stem groups thereof \cite{cano2018mss}. 
These tracks or stems could contain applied audio effects, such as reverberation or dynamic range compression, such that their summation exactly reconstructs the original mixture.

\subsection{Query-based source separation}

Query-based source separators take as input the time-domain signal $\vec{m}$ and a query $\vec{q}_k$ that describes the source $\mathbf{x}_k$ to separate. 
The output is an estimate $\hat{\vec{x}}_k$ of the target source signal:
\begin{equation}
\hat{\mathbf{x}}_k = f_\theta(\mathbf{m}, \mathbf{q}_k).
\end{equation}
Conceptually, an ideal query maximizes mutual information with its target $I(\mathbf{q}_k; \mathbf{x}_k)$ while minimizing it with interfering sources $I(\mathbf{q}_k; \mathbf{x}_{j})$ for all $j \neq k$.
While we do not enforce this criterion explicitly, it serves as a design principle when choosing the query representation.
We define the complete query set $\mathcal{Q} = \{ \mathbf{q}_k \}_{k=1}^K$ and implement the queries $\mathbf{q}_k \in \mathbb{R}^{Q}$ as time-invariant embeddings obtained from a feature extractor $\mathbf{q}_k = \psi(\mathbf{x}_k)$, following prior conventions \cite{watcharasupat:2024:stem-agnostic}.

\subsection{Recursive stem embedding generation}
\label{sec:rec-seg}

In the blind \ac{mss} scenario, both the number of sources $K$ and the query set $\mathcal{Q}$ are unknown \textit{a priori}. We propose estimating the query set recursively with a flow model $u_\phi$ producing one query embedding at a time, conditioned on the mixture and the previously generated queries. We refer to this as \ac{seg}.

Inference is
illustrated in \Cref{fig:diagram}. At iteration $\tilde{k}$, the \ac{seg} model generates a pair $\mathbf{z}_0 = [\mathbf{q}_{\tilde{k}}, \mathbf{r}_{\tilde{k}}]$: the query for the next source $\mathbf{q}_{\tilde{k}}$, and a \emph{residual embedding} $\mathbf{r}_{\tilde{k}}$ describing the sources that remain. The pair is obtained by solving the flow \ac{ode}~\cite{lipman2023flow} from $t=1$ to $t=0$,
\begin{equation}
    \mathrm{d}\mathbf{z}_t = u_\phi\!\left(\mathbf{z}_t, t, \tilde{k}, \mathbf{m},
    \hat{\mathcal{Q}}_{\mathrm{prev}}\right)\mathrm{d}t ,
    \label{eq:ode}
\end{equation}
where $u_\phi$ is a Transformer~\cite{peebles2023scalable} with parameters $\phi$ trained by \ac{cfm}~\cite{lipman2023flow}, and $\hat{\mathcal{Q}}_{\mathrm{prev}} = \{\hat{\mathbf{q}}_k\}_{k<\tilde{k}}$ is the set of queries generated so far.

Generating the residual alongside the query makes the recursion self-terminating. When no sources remain, the model is trained to emit a fixed \emph{stop symbol} $\mathbf{s}$ in the residual slot instead of an embedding of audio. The recursion halts once $\langle \hat{\mathbf{r}}_{\tilde{k}}/\|\hat{\mathbf{r}}_{\tilde{k}}\|, \mathbf{s}\rangle > \tau_{\mathrm{stop}}$.

\begin{algorithm}[t]
\caption{Training of Stem Embedding Generator}
\label{alg:seg-training}
\begin{algorithmic}[1]

\Require feature extractor $\psi$, stop symbol $\mathbf{s}$, parameters $\phi$
\Repeat
    \State Sample $\{\mathbf{x}_1,\dots,\mathbf{x}_K\}$ from dataset in random order
    \State $\mathbf{m} \leftarrow \textstyle\sum_{k=1}^{K}\mathbf{x}_k$ \Comment{Obtain mixture}
    \State $\tilde{k} \sim \mathcal{U}\{1,\dots,K\}$ \Comment{Truncate the recursion}
    \State $\mathbf{q} \leftarrow \psi(\mathbf{x}_{\tilde{k}})$, \quad
           $\mathcal{Q}_\mathrm{prev} \leftarrow \{\psi(\mathbf{x}_k)\}_{k<\tilde{k}}$ 
    \State $\mathbf{r} \leftarrow  \psi(\mathbf{m}-\textstyle\sum_{k=1}^{\tilde{k}}\mathbf{x}_k)$ \quad if \; $\tilde{k} < K$ \; else  \; $\mathbf{s}$
    \State $\mathbf{z}_0 \leftarrow \mathrm{concat}(\mathbf{q}, \mathbf{r})$, \quad
           $\mathbf{z}_1 \sim \mathcal{N}(\mathbf{0},\mathbf{I})$, \quad $t \sim \mathcal{U}[0,1]$
    \State $\mathbf{z}_t \leftarrow t\,\mathbf{z}_1 + (1-t)\,\mathbf{z}_0$ \Comment{Linear interpolant}
    \State $\mathcal{L} \leftarrow \|u_\phi(\mathbf{z}_t;\, t, \tilde{k}, \mathbf{m}, \mathcal{Q}_\mathrm{prev}) 
           - (\mathbf{z}_1 - \mathbf{z}_0)\|^2$ \Comment{CFM loss}
    \State $\phi \leftarrow \mathrm{AdamW}(\phi, \nabla_\phi\mathcal{L})$
\Until{converged}
\end{algorithmic}
\end{algorithm}

Training is described in Algorithm~\ref{alg:seg-training}. 
Each example reconstructs one step of the recursion: we shuffle the stems of a track, draw a depth $\tilde{k}$ uniformly, and ask the model to produce $\psi(\mathbf{x}_{\tilde{k}})$ given the mixture and the descriptors of the preceding stems, with the residual target set to $\mathbf{r}_{\tilde{k}}=\psi(\mathbf{m}-\sum_{k=1}^{\tilde{k}}\mathbf{x}_k)$, or to $\mathbf{r}_{\tilde{k}}=\mathbf{s}$ when $\tilde{k}=K$.
Sampling the depth uniformly means a single pass supervises every stage of the recursion, including its termination, without ever unrolling it: the cost of a training step is independent of $K$. 
Because the stem order is shuffled at training time, no iteration index is associated with any particular instrument, and the order in which queries emerge at inference is expected to be uniformly random.

\section{Experiments}
\label{sec:experiments}

We base our experiments on the MoisesDB dataset \cite{pereira:2023:moisesdb}, using the splits from \cite{watcharasupat:2024:stem-agnostic}.
MoisesDB organizes sources into 11 broad and 38 fine categories, denoted \emph{groups} and \emph{instruments}.
We train all models with both levels, choosing one at random and indicating it with a binary hierarchy-level flag $h$.
The training set is supplemented with 125 songs from MedleyDB \cite{bittner:2014:medleydb} and 10 from AlbumDB \cite{mckenzie:2026:albumdb}, organized to match the MoisesDB taxonomy.
We use wet, processed stems and exclude MoisesDB songs with bleeding tracks.
In total, 279 songs are used for training, 48 for testing, and 26 for validation.

For each training example, we randomly sample a song and extract a 10-s segment, loading its aligned stems from either the group or instrument hierarchy level.
At the \emph{group} level, stems from the same taxonomy group are summed; at the \emph{instrument} level, we merge stems sharing a taxonomy leaf. We always sum all vocal tracks into one unit.
We discard near-silent segments or those with fewer than two active stems based on RMS and relative loudness thresholds, and randomly drop some stems as augmentation.
Each stem is also augmented with random gains and randomized effect chains (filtering, EQ, convolution reverb, compression, etc.); details are available online.\footnote{\href{https://github.com/REMUS-Aalto/Mus3D}{github.com/REMUS-Aalto/Mus3D} (will be available upon acceptance)}

To build the test set, each of the 48 test songs is split into non-overlapping 10-s segments on a fixed grid, giving 1062 segments in total.
Group and instrument-level stems are defined as in training.  
No augmentations are applied.
A source counts as active in a segment, and thus enters the target set $\mathcal{S}$, when its RMS is at least $-50$\,dBFS and no more than 35\,dB below that of the mixture.
Segments contain, on average, 4.5 active groups (1 to 8) or 7.0 instruments (1 to 15).

\subsection{Query descriptor}
\label{sec:experiments-query-descriptor}

As in \cite{watcharasupat:2024:stem-agnostic,watcharasupat:2026:hyperellipsoidal}, the experiments employ the PaSST model \cite{koutini:2022:passt} trained on OpenMIC-2018 \cite{humphrey:2018:openmic}.
The model uses $10$-s long audio clips summed to mono and downsampled to $\qty{32}{\kilo\hertz}$ as input to generate the embeddings.
Additionally, following \cite{watcharasupat:2026:hyperellipsoidal} we apply \ac{pca} using samples from the training datasets to reduce the feature dimension of the embeddings to $128$ with a total explained variance of $0.96$.

\subsection{Stem embedding generator}
\label{sec:experiments-seg}

The SEG architecture is \iac{dit} with 12 layers, a hidden size of 384, and 12 attention heads, totaling roughly 37M parameters.
It is trained on the PCA-reduced 128-dimensional PaSST embeddings, producing one query at a time (see \Cref{fig:diagram}).
The mixture $\mathbf{m}$, encoded both as its PCA-reduced PaSST embedding and as a sequence of DAC-VAE latents \cite{shi:2025:sam} (25\,Hz, 128-dimensional), conditions the \ac{dit} via cross-attention, together with the previously generated queries $\hat{\mathcal{Q}}_{\mathrm{prev}}$.
The flow time $t$, the step counter $\tilde{k}$ and the hierarchy-level flag $h$ condition the model via adaptive layer normalization.
All models are trained on a single NVIDIA H200 GPU.
\ac{seg} is trained for 400k iterations with a batch size of 8.

At inference we solve \eqref{eq:ode} with a stochastic flow sampler \cite{mcallister2026finite}, using $T=32$ uniform steps, and a constant churn parameter $\gamma=0.1$.
The \emph{stop symbol} of \Cref{sec:rec-seg} is the least-variance direction of the \ac{pca}-reduced space, leaving a wide margin to any real source embedding, since this direction carries almost no energy for real sources.
We use a threshold of $\tau_\mathrm{stop}=0.9$, since it did not cause premature stopping for the recursion.
However, as a fallback mechanism, we cap the maximum number of iterations at $16$.

\subsection{Query-based separator models}
\label{sec:experiments-separators}

\noindent \Acfi{ditsep}:
We implement a query-based separator also based on \iac{cfm}-trained \ac{dit}, building upon the \emph{large} version of SAM-Audio \cite{shi:2025:sam}, consisting of roughly 2.2B trainable parameters.\footnote{The original SAM-Audio large implementation has 3B parameters, but we did not use the cross-attention mechanism.}
We adapt the original implementation by conditioning the transformer, via adaptive layer normalization, on the flow time $t$, the 128-dimensional query embeddings, and the hierarchy flag $h$.
We initialize the training with the public weights\footnote{According to \cite{shi:2025:sam}, MoisesDB was not used to train SAM-Audio, so the test set is not seen by the pretrained model.}, and fine-tune it for roughly 300k iterations at a batch size of 16, using our train dataset, tracking the exponential moving average of the weights with a rate of 0.9999.
At inference we use the stochastic flow sampler of \cite{mcallister2026finite} with $T=32$ uniform steps and a constant churn parameter $\gamma=0.5$.

\smallskip \noindent \textit{BS-Locoformer separator}:
We use the musical \ac{bs} variant of the \ac{tf}-Locoformer \cite{saijo:2026:input-adaptive}, in the \emph{medium} setting, with query conditioning via bias modulation of the SwiGLU activations.
Models train for approximately $200$k iterations at a batch size of 4.

\section{Evaluation}
\label{sec:evaluation}

We first assess how accurately \ac{seg} discovers the stems in a mixture (\Cref{sec:eval-discovery}), and then evaluate separation quality with the generated queries against state-of-the-art \ac{mss} baselines using objective (\Cref{sec:eval_separation}) and subjective experiments (\Cref{sec:subjective}).

\subsection{Stem discovery}
\label{sec:eval-discovery}

Although \ac{seg} generates continuous, open-set embeddings, we evaluate stem discovery as a multi-label classification task over the MoisesDB taxonomy, both at \emph{group} and \emph{instrument} levels, allowing a direct comparison with baselines that predict instrument labels.

To assign a label to each generated query, we train a hierarchical stem classifier on the \ac{pca}-reduced PaSST embeddings, following \cite{zhong:2023:attention-based}: a linear layer with sigmoid activations per instrument tag, with group activations obtained by max pooling over their tags, trained with a weighted binary cross-entropy loss at both levels on individual stems and group sums.
Each query is assigned the highest-activation tag.
Since the evaluation is bounded by the classifier accuracy, we include an Oracle classifier operating directly on \ac{gt} stem embeddings to provide an upper bound for \ac{seg}.

We compare \ac{seg} against two baselines that predict the stems directly from the mixture.
The first is a multi-label classifier with the same architecture and loss as the stem classifier, trained on the PCA-reduced PaSST embeddings of the \ac{seg} training mixtures to predict all their active tags.
It outputs every tag whose activation exceeds a tag-specific threshold, chosen to maximize the $\mathrm{F}_1$ of that tag on the validation set.
The second is MusicFlamingo \cite{ghosh2026music}, a language model for music understanding, which we prompt to list the stems in a mixture.\footnote{Group level: ``\textit{Which instruments and voices can you hear in this recording? Answer with a comma-separated list.}'' Instrument level: ``\textit{List every instrument and voice you can hear as specifically as possible. Break a drum kit into its separable pieces. Answer with a comma-separated list.}''}
The resulting phrases are mapped onto the taxonomy with a deterministic lexical procedure provided in our code.
We also evaluate a variant of \ac{seg} that is additionally conditioned on the residual embedding generated at the previous iteration, which summarizes the stems yet to be extracted.

For each $10$-s test segment, we compare the predicted labels $\hat{\mathcal{S}}$, with one per claimed stem, against the active stem labels $\mathcal{S}$, computing precision $\mathrm{P}=|\mathcal{S} \cap \hat{\mathcal{S}}|/|\hat{\mathcal{S}}|$, recall $\mathrm{R}=|\mathcal{S} \cap \hat{\mathcal{S}}|/|\mathcal{S}|$, $\mathrm{F}_1 = 2\mathrm{P}\mathrm{R}/(\mathrm{P}+\mathrm{R})$, and the count error $\Delta K=\bigl||\hat{\mathcal{S}}|-|\mathcal{S}|\bigr|$, 
which reflects the difference in number of predicted stems, independent of their labels. 
All metrics are averaged over segments in \Cref{tab:discovery}.

\begin{table}[]
\vspace{-6.2pt}
\caption{
Objective evaluation of stem discovery ,
scored per 10-s segment and aggregated over the 1062 test segments (mean $\pm$ std.). The best result in each column is highlighted in bold.
}
\label{tab:discovery}
\centering
\resizebox{\columnwidth}{!}{\begin{tabular}{@{}ll|cccc@{}}
\toprule
& \textbf{Method} & \textbf{F$_1$} & \textbf{Precision} & \textbf{Recall} & $\Delta K$ \\ \midrule \midrule
\multirow{5}{*}{\shortstack[l]{\textit{All}\\\textit{Groups}}}
& Oracle Classifier & 0.96{\scriptsize$\,\pm\,0.10$} & 0.96{\scriptsize$\,\pm\,0.10$} & 0.96{\scriptsize$\,\pm\,0.10$} & 0.00{\scriptsize$\,\pm\,0.00$} \\
\cmidrule{2-6}
& Multi-label Classifier & \textbf{0.89}{\scriptsize$\,\pm\,0.12$} & 0.88{\scriptsize$\,\pm\,0.15$} & \textbf{0.91}{\scriptsize$\,\pm\,0.14$} & 0.69{\scriptsize$\,\pm\,0.82$} \\
& MusicFlamingo & 0.82{\scriptsize$\,\pm\,0.19$} & 0.87{\scriptsize$\,\pm\,0.21$} & 0.80{\scriptsize$\,\pm\,0.22$} & 1.04{\scriptsize$\,\pm\,1.07$} \\
& SEG & 0.88{\scriptsize$\,\pm\,0.13$} & \textbf{0.92}{\scriptsize$\,\pm\,0.14$} & 0.86{\scriptsize$\,\pm\,0.16$} & \textbf{0.67}{\scriptsize$\,\pm\,0.78$} \\
& SEG (with res. embedding) & 0.85{\scriptsize$\,\pm\,0.14$} & 0.87{\scriptsize$\,\pm\,0.16$} & 0.85{\scriptsize$\,\pm\,0.16$} & 0.71{\scriptsize$\,\pm\,0.81$} \\ \midrule \midrule
\multirow{5}{*}{\shortstack[l]{\textit{All}\\\textit{Instr.}}}
& Oracle Classifier & 0.86{\scriptsize$\,\pm\,0.16$} & 0.86{\scriptsize$\,\pm\,0.16$} & 0.86{\scriptsize$\,\pm\,0.16$} & 0.00{\scriptsize$\,\pm\,0.00$} \\
\cmidrule{2-6}
& Multi-label Classifier & \textbf{0.75}{\scriptsize$\,\pm\,0.16$} & 0.70{\scriptsize$\,\pm\,0.19$} & \textbf{0.84}{\scriptsize$\,\pm\,0.17$} & 2.01{\scriptsize$\,\pm\,1.80$} \\
& MusicFlamingo & 0.33{\scriptsize$\,\pm\,0.14$} & 0.44{\scriptsize$\,\pm\,0.19$} & 0.27{\scriptsize$\,\pm\,0.14$} & 3.17{\scriptsize$\,\pm\,2.21$} \\
& SEG & 0.74{\scriptsize$\,\pm\,0.18$} & \textbf{0.77}{\scriptsize$\,\pm\,0.20$} & 0.72{\scriptsize$\,\pm\,0.19$} & \textbf{1.26}{\scriptsize$\,\pm\,1.18$} \\
& SEG (with res. embedding) & 0.70{\scriptsize$\,\pm\,0.18$} & 0.69{\scriptsize$\,\pm\,0.20$} & 0.73{\scriptsize$\,\pm\,0.19$} & 1.34{\scriptsize$\,\pm\,1.30$} \\ \bottomrule
\vspace{-15pt}
\end{tabular}}
\end{table}

As shown in \Cref{tab:discovery}, \ac{seg} achieves the highest precision and the lowest $\Delta K$ at both levels, with an $\mathrm{F}_1$ on par with the multi-label classifier ($0.88$ vs.\ $0.89$ for groups, and $0.74$vs.\ $0.75$ for instruments).
MusicFlamingo is competitive at the group level ($\mathrm{F}_1=0.82$), but degrades sharply at the instrument level ($\mathrm{F}_1=0.33$), suggesting that fine-grained stem identification remains challenging for language models.
The oracle reaches $\mathrm{F}_1$ of $0.96$ (groups) and $0.86$ (instruments), showing part of the error stems from the classifier.
Conditioning on the previous residual embedding degrades all metrics for \ac{seg}, which we attribute to exposure bias: the model is trained on \ac{gt} residual embeddings but receives its own estimates at inference.

\subsection{Source separation objective evaluation}
\label{sec:eval_separation}

\begin{table}[]
\vspace{-6.2pt}
\caption{Objective metrics for source separation performance. Scored only on sources detected by all blind methods (mean $\pm$ std.).
Best result per block in bold.
}
\label{tab:separation}

\resizebox{\columnwidth}{!}{\centering
\begin{tabular}{@{}lll|lcc@{}}
\toprule
& \textbf{Setting} & \textbf{Separator} & \textbf{Judge} $\uparrow$ & \textbf{SNR (dB)} $\uparrow$ & \textbf{MERT-MSE} $\downarrow$ \\ \midrule \midrule
\multirow{13}{*}{\shortstack[l]{\textit{Groups:}\\\textit{VDBGP}}}
& \multirow{4}{*}{Reference} & Mixture & 1.2{\scriptsize$\,\pm\,0.41$} & -4.9{\scriptsize$\,\pm\,8.3$} & 0.22{\scriptsize$\,\pm\,0.05$} \\
& & Ground truth & 4.7{\scriptsize$\,\pm\,0.47$} & $\infty$ & 0 \\
& & Ground truth (AE) & 4.7{\scriptsize$\,\pm\,0.47$} & 14{\scriptsize$\,\pm\,9.6$} & 0.04{\scriptsize$\,\pm\,0.02$} \\
& & Ideal Ratio Mask & 4.4{\scriptsize$\,\pm\,0.58$} & 10{\scriptsize$\,\pm\,6.6$} & 0.07{\scriptsize$\,\pm\,0.03$} \\ \cmidrule{2-6}
& \multirow{2}{*}{Static} & HTDemucs & 4.4{\scriptsize$\,\pm\,0.67$} & \textbf{9.7}{\scriptsize$\,\pm\,5.0$} & 0.12{\scriptsize$\,\pm\,0.05$} \\
& & BS-Locoformer & 4.1{\scriptsize$\,\pm\,0.80$} & 8.0{\scriptsize$\,\pm\,4.0$} & 0.14{\scriptsize$\,\pm\,0.05$} \\ \cmidrule{2-6}
& \multirow{1}{*}{Oracle text} & SAM-A.  & 4.4{\scriptsize$\,\pm\,1.01$} & 4.0{\scriptsize$\,\pm\,6.7$} & 0.15{\scriptsize$\,\pm\,0.06$} \\ \cmidrule{2-6}
&  \multirow{2}{*}{Oracle emb.} & BS-Locoformer & 4.3{\scriptsize$\,\pm\,0.69$} & 8.4{\scriptsize$\,\pm\,4.8$} & 0.13{\scriptsize$\,\pm\,0.05$} \\
&  & DiT-Sep & \textbf{4.7}{\scriptsize$\,\pm\,0.49$} & 6.6{\scriptsize$\,\pm\,5.3$} & \textbf{0.11}{\scriptsize$\,\pm\,0.05$} \\ \cmidrule{2-6}
& Other song emb. & Banquet  & 4.1{\scriptsize$\,\pm\,1.04$} & 7.6{\scriptsize$\,\pm\,7.1$} & 0.14{\scriptsize$\,\pm\,0.06$} \\ \cmidrule{2-6}
&  \multirow{3}{*}{Blind}& MusicFlam. + SAM-A. & 4.2{\scriptsize$\,\pm\,1.24$} & 2.7{\scriptsize$\,\pm\,7.1$} & 0.16{\scriptsize$\,\pm\,0.07$} \\
& & MuS3D (BS-Locoformer)  & 4.3{\scriptsize$\,\pm\,0.70$} & 8.1{\scriptsize$\,\pm\,4.7$} & 0.13{\scriptsize$\,\pm\,0.05$} \\
& & MuS3D (DiT-Sep) & \textbf{4.7}{\scriptsize$\,\pm\,0.45$} & 6.5{\scriptsize$\,\pm\,5.1$} & 0.12{\scriptsize$\,\pm\,0.05$} \\ \midrule \midrule
\multirow{9}{*}{\shortstack[l]{\textit{All}\\\textit{Instr.}}}
& \multirow{4}{*}{Reference} & Mixture & 1.3{\scriptsize$\,\pm\,0.41$} & -9.1{\scriptsize$\,\pm\,8.3$} & 0.25{\scriptsize$\,\pm\,0.05$} \\
& & Ground truth & 4.5{\scriptsize$\,\pm\,0.59$} & $\infty$ & 0 \\
& & Ground truth (AE) & 4.6{\scriptsize$\,\pm\,0.54$} & 13{\scriptsize$\,\pm\,10$} & 0.04{\scriptsize$\,\pm\,0.02$} \\
& & Ideal Ratio Mask & 4.2{\scriptsize$\,\pm\,0.76$} & 9.1{\scriptsize$\,\pm\,5.4$} & 0.10{\scriptsize$\,\pm\,0.05$} \\ \cmidrule{2-6}
& \multirow{1}{*}{Oracle text} & SAM-A.  & 4.3{\scriptsize$\,\pm\,1.18$} & -2.4{\scriptsize$\,\pm\,8.9$} & 0.21{\scriptsize$\,\pm\,0.09$} \\ \cmidrule{2-6}
& \multirow{2}{*}{Oracle emb.}& BS-Locoformer & 4.1{\scriptsize$\,\pm\,0.84$} & \textbf{7.5}{\scriptsize$\,\pm\,4.9$} & 0.15{\scriptsize$\,\pm\,0.06$} \\
& & DiT-Sep & \textbf{4.6}{\scriptsize$\,\pm\,0.56$} & 6.1{\scriptsize$\,\pm\,6.4$} & \textbf{0.12}{\scriptsize$\,\pm\,0.06$} \\ \cmidrule{2-6}
& \multirow{2}{*}{Blind} & MuS3D (BS-Locoformer)  & 4.1{\scriptsize$\,\pm\,0.83$} & 6.8{\scriptsize$\,\pm\,5.5$} & 0.16{\scriptsize$\,\pm\,0.06$} \\
& & MuS3D (DiT-Sep) & \textbf{4.6}{\scriptsize$\,\pm\,0.56$} & 5.6{\scriptsize$\,\pm\,6.6$} & 0.14{\scriptsize$\,\pm\,0.07$} \\ \bottomrule
\vspace{-15pt}
\end{tabular}}
\end{table}

Having assessed which stems \ac{seg} discovers, we now evaluate the separation performance when using generated queries.
Each separated stem is scored against the \ac{gt} stem matching the label assigned by the stem classifier.
To avoid comparing mislabeled outputs, we restrict the evaluation to stems detected by \ac{seg} and, at the group level, also by MusicFlamingo.
At the group level, we further restrict the evaluation to \ac{vdbgp}, to compare with static separators trained on these stems.
This leaves us with 3422 group stems and 5197 individual instruments.

We compare against two static 6-stem separators: the publicly available HTDemucs \cite{rouard2023hybrid}, and a BS-Locoformer trained on \ac{vdbgpo} with our training data.
As query-based baselines, we evaluate Banquet \cite{watcharasupat:2024:stem-agnostic} with queries taken from a different song containing the same stem, following its original evaluation protocol, and SAM-Audio \cite{shi:2025:sam} with text prompts given either by the taxonomy class names (oracle) or, only at the group level, by the phrases produced by MusicFlamingo in \Cref{sec:eval-discovery} (blind).
Our BS-Locoformer and \ac{ditsep} separators are evaluated both with oracle queries extracted from the \ac{gt} stems and with the queries generated by \ac{seg}, the latter, corresponding to the proposed \ac{ms3d}.
As references, we report the mixture, the ideal ratio mask \cite{vincent2007oracle}, and the \ac{gt} stem after encoding and decoding with the DAC-VAE autoencoder used in the \ac{ditsep}, Ground truth (AE).

We measure separation quality with three metrics, all computed on mono signals.
The SAM-Audio Judge \cite{wang2026samjudge} is a learned, reference-free model that predicts perceptual quality scores for a separated stem given the mixture and a text prompt as context, for which we use the taxonomy label of the target; we report its ``Overall'' score.
We report the \ac{snr}, which compares waveforms and therefore penalizes imperceptible phase differences, such as those introduced by the autoencoder of the \ac{ditsep}.
We also report the mean-squared error between MERT representations \cite{li2024mert} (12th layer, following \cite{bereuter2025towards}), a perceptually motivated metric that correlates with subjective DMOS ratings and is insensitive to slight phase differences; we refer to it as MERT-MSE.

As shown in \Cref{tab:separation}, both separators (BS-Locoformer, \ac{ditsep}) perform almost equally well with \ac{seg} queries and oracle queries at both levels: the Judge score is unchanged, and MERT-MSE increases by at most $0.01$, so the separators are robust to estimation errors from \ac{seg}.
This also holds at the instrument level, where \ac{seg} generates more numerous, finer-grained queries, and \ac{ms3d} matches the group-level scores.
Remarkably, \ac{ms3d} (\ac{ditsep}) obtains the highest Judge score at both levels (4.7 and 4.6), on par with \ac{gt} stems.

SAM-Audio scores lower, even with oracle text prompts, and degrades further with MusicFlamingo prompts, often extracting the wrong stem.
HTDemucs obtains the highest SNR and a MERT-MSE comparable to the best query-based separator, but a lower Judge score than the \ac{ditsep}.
The \ac{ditsep} obtains a comparatively low SNR as it generates waveforms from latents; the higher SNR of autoencoded \ac{gt} stems ($14$ and $13$\,dB) shows the autoencoder is partly responsible.

\subsection{Source separation subjective evaluation}
\label{sec:subjective}

\definecolor{wincolor}{HTML}{007b25} \definecolor{loscolor}{HTML}{c5283d} 
\begin{figure}[t]
\centering
\resizebox{\columnwidth}{!}{\footnotesize
\begin{tikzpicture}[x=1cm, y=1cm]
  \def\W{4.0}    \def\H{0.38}   \def\dy{0.52}  \def\gap{0.10} \foreach \a/\b/\wa/\wb/\p/\s/\g [count=\i] in {
      {MuS3D (DiT-Sep)}/{HTDemucs}/116/14/{$p<0.001$}/1/1,
      {MuS3D (DiT-Sep)}/{MusicFlam. + SAM-A.}/109/21/{$p<0.001$}/1/1,
      {MuS3D (DiT-Sep)}/{Oracle (DiT-Sep)}/59/71/{$p=0.34$}/0/2,
      {MuS3D (DiT-Sep)}/{Ground truth (AE)}/61/69/{$p=0.54$}/0/2,
      {MuS3D (DiT-Sep)}/{Ground truth}/44/86/{$p<0.001$}/2/2,
      {Oracle (DiT-Sep)}/{Ground truth (AE)}/68/62/{$p=0.66$}/0/3,
      {Oracle (DiT-Sep)}/{Ground truth}/41/89/{$p<0.001$}/2/3}{
    \pgfmathsetmacro{\y}{-(\i-1)*\dy-(\g-1)*\gap}
    \pgfmathsetmacro{\xs}{\W*\wa/(\wa+\wb)}
    \ifcase\s
      \def\ca{gray!45}\def\cb{gray!25}\def\ta{black}\def\tb{black}\def\fa{}\def\fb{}
    \or
      \def\ca{wincolor!75!black}\def\cb{loscolor!70}\def\ta{white}\def\tb{white}\def\fa{\bfseries}\def\fb{}
    \or
      \def\ca{loscolor!70}\def\cb{wincolor!75!black}\def\ta{white}\def\tb{white}\def\fa{}\def\fb{\bfseries}
    \fi
    \fill[\ca] (0, \y-\H/2) rectangle (\xs, \y+\H/2);
    \fill[\cb] (\xs, \y-\H/2) rectangle (\W, \y+\H/2);
    \draw[white, line width=0.6pt] (\xs, \y-\H/2) -- (\xs, \y+\H/2);
    \node[anchor=west, text=\ta, font=\fa, inner sep=2pt] at (0, \y) {\wa};
    \node[anchor=east, text=\tb, font=\fb, inner sep=2pt] at (\W, \y) {\wb};
    \node[anchor=east, font=\fa] at (-0.1, \y) {\a};
    \node[anchor=west, font=\fb] at (\W+0.1, \y) {\b};
    \node[anchor=west] at (\W+2.9, \y) {\p};
  }
  \pgfmathsetmacro{\ytop}{0.5*\dy-0.04}
  \pgfmathsetmacro{\ybot}{-6.5*\dy-2*\gap}
  \draw[black!90, densely dashed] (\W/2, \ytop) -- (\W/2, \ybot);
  \draw[semithick] (current bounding box.west |- 0, \ytop) -- (current bounding box.east |- 0, \ytop);
  \foreach \ysep in {-1.5*\dy-0.5*\gap, -4.5*\dy-1.5*\gap}
    \draw[black!35, very thin] (current bounding box.west |- 0, \ysep) -- (current bounding box.east |- 0, \ysep);
  \draw[semithick] (current bounding box.west |- 0, \ybot) -- (current bounding box.east |- 0, \ybot);
\end{tikzpicture}
}
\vspace{-18pt}
\caption{Pairwise preference listening test results from 130 trials per pair. Green/red indicate a significant preference ($p<0.05$), and the winning method is highlighted in bold in such cases.}
\label{fig:preference_results}
\end{figure}

For the subjective evaluation, we follow the pairwise comparison experiment setup used in \cite{moliner:2025:automatic}.
Trials use 4-s excerpts from the MoisesDB test set.
We restrict the taxonomy to \ac{vdbgp}, since this performed the most consistently across the models.
Each trial presents the participant with the song mixture and two examples of separated stems.
The task is to pick the option that better separates the indicated source, i.e., vocals, bass, drums, guitar, or piano.
The listening test involved 13 participants with no reported hearing problems and with prior listening test experience, conducted in soundproof booths of the Aalto Acoustics Lab using headphones at constant loudness.
The test used 2 songs per instrument, shared across all compared pairs to avoid selection bias.
SAM-Audio was prompted with the phrases produced by MusicFlamingo, and all excerpts were taken from stems detected by both MusicFlamingo and \ac{seg}, so that separation errors are attributable to the separator.

\Cref{fig:preference_results} shows the results for the subjective listening test.
For each pair, we test whether the number of times a condition was chosen differs from random chance with a two-sided binomial test, treating $p<0.05$ as significant.
The comparisons show that \ac{ms3d} (\ac{ditsep}) significantly outperforms the baselines (HTDemucs and SAM-Audio), achieves parity with Oracle (\ac{ditsep}) and Ground truth (AE), but is significantly outperformed by the original, unencoded stem.
Oracle (\ac{ditsep}) and Ground truth (AE) are not significantly different, but both are outperformed by the Ground truth.
This indicates that the limiting factor is the autoencoder, not the separation or query generation method.

\section{Conclusion}
\label{sec:conlusion}

This paper proposes \ac{ms3d}, a framework for blind music source separation that automatically discovers the stems in a mixture using audio-query generation.
Its \ac{seg} module recursively generates one embedding per stem, conditioned on the mixture and on the previously generated embeddings of the other stems.
The embeddings can then condition a query-based separator, so no text or audio example has to be supplied by the user.

The \ac{seg} outperforms a language-model baseline at identifying the sources in a mixture.
On correctly detected sources, separators conditioned on the generated queries perform on par with those conditioned on \ac{gt} queries, in both objective metrics and a formal listening test.
The listening test results also suggest that, for the \ac{ditsep} separator, the audio autoencoder is the primary limiting factor on perceived quality. 
This paper shows that audio queries can be generated automatically without loss in separation quality, enabling \ac{mss} when the stems in a mixture are not known in advance.

That said, we identify limitations that should be addressed in future work. \ac{seg} still misclassifies or misses some stems, which lowers recall (\Cref{tab:discovery}).
The evaluation is limited to 10-s segments; extending it to whole songs, where queries must stay consistent across segments, remains open.
Finally, the inference pipeline, which samples from \ac{seg} and then runs \ac{ditsep}, is computationally expensive, but techniques such as flow maps \cite{boffi2026build} offer a route for acceleration.

\section{Acknowledgments}
This research was funded by the Research Council of Finland (grant no. 371845). We acknowledge the computational resources provided by the Aalto Science-IT project. This work was supported by the HUCE infrastructure of the Aalto School of Electrical Engineering. 
Claude (using Opus 5 hosted by Anthropic) and Mistral Vibe (using Mistral Medium 3.5 and Z.ai GLM-5.3, both hosted by Mistral AI) were used to assist with software development, and to draft and revise parts of the manuscript text.
All content was reviewed, edited, and verified by the authors, who take full responsibility for the paper.

\renewcommand{\thebibliography}[1]{\section{References}
\vspace{-5pt}
  \list
  {[\arabic{enumi}]}{\settowidth\labelwidth{[#1]}\leftmargin\labelwidth
   \advance\leftmargin\labelsep
   \usecounter{enumi}\setlength{\itemsep}{0.7pt plus 0.2ex}\setlength{\parsep}{0pt plus 0.2ex}\setlength{\topsep}{0pt plus 0.2ex}}\def\newblock{\hskip .11em plus .33em minus .07em}
  \sloppy\clubpenalty4000\widowpenalty4000
  \sfcode`\.=1000\relax}
\let\endthebibliography=\endlist
\bibliographystyle{IEEEbib}
\bibliography{refs}

\end{document}